\documentclass[aps,prl,twocolumn,superscriptaddress,showpacs,amsmath,amssymb]{revtex4-2}
\usepackage{amsmath}
\usepackage{amsfonts}
\usepackage{amssymb}
\usepackage{graphicx}
\usepackage{color}

\usepackage{bm}
\usepackage{braket}
\usepackage{hyperref}
\usepackage[boxsize=0.25em,centertableaux]{ytableau}

\begin{document}

\title{Gapped topological spin-orbital liquid on the honeycomb lattice}

\author{Masahiko G. Yamada}
\email{masahiko.yamada@aoni.waseda.jp}
\affiliation{Waseda Institute for Advanced Study, Waseda University, Shinjuku-ku, Tokyo 169-0051, Japan.}
\affiliation{Department of Physics, School of Science, the University of Tokyo, Hongo, Bunkyo-ku 113-0033, Japan.}
\date{\today}

\begin{abstract}
We perform large-scale density matrix renormalization group simulations of
the $\mathrm{SU}(4)$ Heisenberg model on the honeycomb lattice
to address the long-standing question of its ground state in an unbiased
and quantitatively controlled manner.  We find reliable numerical evidence
that the ground state is a gapped spin-orbital liquid, presumably with a $Z_4$
topological order, characterized by a finite topological entanglement entropy
close to $\ln(4)$, the absence of both $\mathrm{SU}(4)$ and lattice symmetry breaking,
and a variationally optimized ground-state energy well below
the previously proposed $\pi$-flux variational state.  By exploiting full $\mathrm{SU}(4)$
symmetry and keeping up to 12,800 $\mathrm{SU}(4)$ multiplets, corresponding
to more than one million $\mathrm{U}(1)$ states, we achieve unprecedented accuracy
for two-dimensional $\mathrm{SU}(4)$ quantum magnets.  Finite-size scaling
of energies and entanglement entropies supports a robust gapped phase
in the two-dimensional limit, while a gapless critical state on narrow cylinders
is identified as a proximate remnant of a Dirac spin-orbital liquid.
Our results find the $\mathrm{SU}(4)$ honeycomb Heisenberg model
a realization of a gapped topological spin-orbital liquid and
provide convincing numerical evidence for topological order
in a highly symmetric two-dimensional quantum magnet.
\end{abstract}

\maketitle

\textit{Introduction}. ---
The nature of the ground state of the $\mathrm{SU}(4)$ Heisenberg model on the honeycomb lattice
has remained a long-standing open problem in the study of quantum spin-orbital liquids~\cite{Savary2017,Corboz2012}.
Previous variational and field-theoretical studies have suggested competing gapless
Dirac spin-orbital liquid and gapped spin-orbital liquid states~\cite{Corboz2012,Calvera2021},
but unbiased large-scale numerical evidence capable of distinguishing these scenarios has been lacking.
As a result, whether this model hosts a gapped topological phase in the two-dimensional (2D)
limit has remained unresolved.

Here we provide unbiased numerical evidence toward resolving this issue
by performing state-of-the-art density matrix renormalization
group (DMRG) simulations~\cite{White1992} with full $\mathrm{SU}(4)$ symmetry~\cite{McCulloch2002,Nataf2018,Chen2021}
for the honeycomb lattice.  We provide numerical evidence that the
ground state is a gapped spin-orbital liquid~\cite{Yamada2022}, presumably with a $Z_4$ topological order,
characterized by (i) a finite topological entanglement entropy close to $\ln(4)$,
(ii) the absence of both $\mathrm{SU}(4)$ and lattice symmetry breaking as revealed by entanglement
spectra and real-space observables, and (iii) a highly accurate variational energy
well below the previously proposed $\pi$-flux state.  These results identify
the $\mathrm{SU}(4)$ honeycomb Heisenberg model as a leading microscopic
candidate for a gapped spin-orbital liquid.

The $\mathrm{SU}(4)$ Heisenberg model on the honeycomb lattice is not merely a theoretical
toy model, but is directly relevant to cold atomic systems with emergent $\mathrm{SU}(N_c)$
symmetry, where $N_c$ is the number of colors~\cite{Cazalilla2014}, and
to solid-state realizations of spin-orbital physics, such as
$\alpha$-ZrCl$_3$~\cite{Yamada2018,Yamada2021}.  Resolving its ground-state
phase provides an important benchmark for both quantum simulation
platforms and candidate spin-orbital liquid materials.

Recent theoretical work has proposed a Dirac spin-orbital liquid as a proximate or
unstable phase in $\mathrm{SU}(4)$ honeycomb systems~\cite{Calvera2021}.
Our identification of a gapless critical state on narrow cylinders and its evolution
toward a gapped topological phase provides a direct numerical
realization of this scenario, and clarifies the fate of the Dirac spin-orbital
liquid in an unbiased setting.

To achieve this level of accuracy, we exploit a full non-Abelian $\mathrm{SU}(4)$ symmetry
implementation in DMRG and keep up to 12,800 $\mathrm{SU}(4)$ multiplets, corresponding to
more than one million $\mathrm{U}(1)$ states.  This enables unprecedented accuracy for
2D $\mathrm{SU}(4)$ quantum magnets and reliable access
to entanglement properties and finite-size scaling toward the 2D limit.

Our results address a central question in frustrated quantum magnetism:
whether a simple, symmetry-preserving 2D microscopic Hamiltonian can
realize intrinsic topological order without fine tuning.  The identification of a gapped
spin-orbital liquid in the $\mathrm{SU}(4)$ Heisenberg model
on the honeycomb lattice provides a concrete route from
$\mathrm{SU}(N_c)$-symmetric physics to entangled states,
and connects problems in frustrated magnetism, topological order, and
cold-atom $\mathrm{SU}(N_c)$ systems.

\textit{Methods}. ---
Following Nataf and Mila~\cite{Nataf2018}, we group an $\mathrm{SU}(N_c)$
multiplet by an irreducible representation (irrep) of $\mathrm{SU}(N_c)$.
An irrep is associated with a Young tableau.  For example,
when $N_c=4$, all irreps of $\mathrm{SU}(4)$ are labeled by
a Young tableau $\alpha = (\alpha_1,\,\alpha_2,\,\alpha_3,\,\alpha_4)$,
where $\alpha_i$ is the length of the $i$th row.  A Young tableau
$(\alpha_1,\,\alpha_2,\,\alpha_3,\,\alpha_4)$ is identified with
$(\alpha_1+l,\,\alpha_2+l,\,\alpha_3+l,\,\alpha_4+l)$ for some
$l\in\mathbb{Z}$, so we conventionally set $\alpha_4=0$.  An adjoint
representation \ydiagram{2,1,1}~will be called (2, 1, 1, 0).
We note that for each irrep $\alpha$, the quadratic Casimir $C_2$
of $\alpha$ is defined as
\begin{equation}
    C_2 = \frac{1}{2N_c}\left[n\left(N_c-\frac{n}{N_c}\right) + \sum_{i=1}^{N_c} \alpha_i^2 - \sum_{j=1}^{\alpha_1} c_j^2 \right],
\end{equation}
where $c_j$ ($j=1,\dots,\alpha_1$) are the lengths of the columns.

One can define $9\nu$ coefficients by Clebsch-Gordan coefficients (CGCs)
as follows~\cite{Chen2002}.
\begin{widetext}
\begin{equation}
\begin{bmatrix}
\nu_1 & \nu_2 & \nu_{12} \\
\nu_3 & \nu_4 & \nu_{34} \\
\nu_{13} & \nu_{24} & \nu
\end{bmatrix}_{\tau_{13}\tau_{24}\tau^\prime}^{\tau_{12}\tau_{34}\tau}
=\sum_{\{m\}}
{C^*}_{m_1 m_3 m_{13}}^{\nu_1\nu_3\nu_{13};\tau_{13}}
{C^*}_{m_2 m_4 m_{24}}^{\nu_2\nu_4\nu_{24};\tau_{24}}
{C^*}_{m_{13} m_{24} 1}^{\nu_{13}\nu_{24}\nu;\tau^\prime}
C_{m_1 m_2 m_{12}}^{\nu_1\nu_2\nu_{12};\tau_{12}}
C_{m_3 m_4 m_{34}}^{\nu_3\nu_4\nu_{34};\tau_{34}}
C_{m_{12} m_{34} 1}^{\nu_{12}\nu_{34}\nu;\tau},
\end{equation}
\end{widetext}
where $C_{m_1 m_2 m_{12}}^{\nu_1\nu_2\nu_{12};\tau_{12}}$ is
a CGC of $\mathrm{SU}(N_c)$ when irreps $\nu_1$ and $\nu_2$
are combined into $\nu_{12}$ with an outer multiplicity $\tau_{12}$.
However, this formula is not that useful for the actual calculation.
It is more practical to employ the Schur-Weyl duality and to rewrite
the expression by subduction coefficients (SDCs) of symmetric
groups~\cite{Nataf2018,Chen2002}.  We note that our method based
on the Schur-Weyl duality can be generalized from $\mathrm{SU}(N_c)$
to any classical groups.

One has to calculate four types of $9\nu$ coefficients for the DMRG simulation
for the ground state.
\begin{equation}
\begin{bmatrix}
\alpha & \ydiagram{1} & \beta \\
\ydiagram{2,1,1} & \cdot & \ydiagram{2,1,1} \\
\alpha' & \ydiagram{1} & \beta'
\end{bmatrix}
\begin{bmatrix}
\alpha & \ydiagram{1} & \beta \\
\cdot & \ydiagram{2,1,1} & \ydiagram{2,1,1} \\
\alpha & \ydiagram{1} & \beta'
\end{bmatrix}
\begin{bmatrix}
\alpha & \ydiagram{1} & \beta \\
\ydiagram{2,1,1} & \ydiagram{2,1,1} & \cdot \\
\alpha' & \ydiagram{1} & \beta
\end{bmatrix}
\begin{bmatrix}
\alpha & \beta & \gamma \\
\ydiagram{2,1,1} & \ydiagram{2,1,1} & \cdot \\
\alpha' & \beta' & \gamma
\end{bmatrix},
\end{equation}
where $\alpha$, $\beta$, $\alpha'$ and $\beta'$ are all relevant irreps used
in the calculation, $\gamma$ represents the ground state symmetry sector,
and $\cdot$ represents the trivial representation.
The first three types can be computed easily, while the last requires efforts.
Thus, we calculate all the relevant coefficients once and for all before
the simulation, and store them in the hash table.  We note that one need not
exhaust all of the coefficients about the last type and that about half of them
are enough for the simulation due to the symmetry of $9\nu$ coefficients.
Details are included in Supplemental Material~\cite{SM}.  In addition, the eigenstate
prediction which accelerates the Lanczos iteration requires additional $3\nu$ and $6\nu$
(Racah) coefficients~\cite{Sharma2012}.  Details of the implementation of
the eigenstate prediction will be discussed in the future publication.

Although our method is overlapping with the work by Nataf and Mila~\cite{Nataf2018},
we use a truncation scheme different from Nataf and Mila's.  First,
the definition of a ``bond dimension'' $m$ is different, and we use an ordinary definition
where $m$ denotes the number of states before enlarging the system
or environment block.  Second, we keep irreps with a width $\alpha_1$
until $\alpha_1 \leq \alpha_\mathrm{max}$.  For $\mathrm{SU}(4)$, we use
$\alpha_\mathrm{max}=9$, corresponding to taking $M=220$ irreps
from the smallest $\alpha_1$.  For $\mathrm{SU}(3)$, we use $M=105$.
Extrapolating the $M \to \infty$ limit is very difficult as the computational
cost grows rapidly.  Currently we do not even know whether it grows
exponentially about $M$ or not.  We have checked that $M=220$ is sufficient
until $m=12800$ and $L_y=12$.

\textit{Simulations}. ---
The $\mathrm{SU}(4)$ Heisenberg model on the honeycomb lattice
is defined in two ways as follows.
\begin{equation}
    H = \sum_{\langle ij \rangle} \left(2\bm{S}_i\cdot \bm{S}_j+\frac{1}{2}\right)\left(2\bm{T}_i\cdot \bm{T}_j+\frac{1}{2}\right) = \sum_{\langle ij \rangle} P_{ij},
\end{equation}
where $\bm{S}_i$ are spin-1/2 operators for the spin sector,
$\bm{T}_i$ are spin-1/2 operators for the orbital sector,
and $P_{ij}$ is a swapping operator between two fundamental
representations on the $i$th and $j$th sites of the honeycomb lattice.
Here a fundamental representation is decomposed into the
spin and orbital sectors by a simple tensor product.
$\langle ij \rangle$ runs over every nearest-neighbor bond of
the honeycomb lattice.

\begin{figure}
    \centering
    \includegraphics[width=8.6cm]{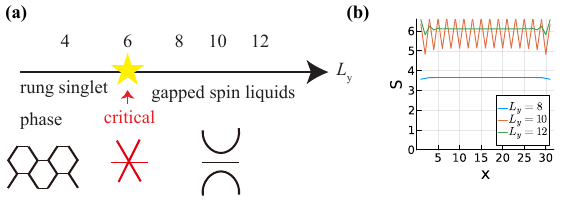}
    \caption{(a) Phase diagram about $L_y$ (cylinder DMRG)
    with a small figure of its phase.  The rung
    singlet phase appears when $L_y=4$, while a gapped spin liquid phase appears
    when $L_y \geq 8$.  Transition occurs at $L_y=6$, which is gapless and
    belongs to the $\mathrm{SU}(4)$ level-1 Wess-Zumino-Witten universality class,
    consistent with a quasi-one-dimensional remnant of a Dirac spin-orbital liquid.
    (b) Entanglement entropy $S$ observed by bipartition of the cylinder
    disconnected around the cross section at $x$.  Approximate independence of
    $x$ shows the gapped nature of the phase.  Even-odd effect only appears in
    the $L_y=10$ case.}
    \label{ov}
\end{figure}

The geometry of the honeycomb lattice is always a cylinder geometry
with $L_y$ being the number of sites, not unit cells, around the
circumference.  We always use a zigzag edge boundary condition,
and $L_x$ is the number of zigzag chains along the cylinder.
This means that we have $N = L_x L_y$ sites in total.
$(L_x,\,L_y)=(n,\,2m)$ corresponds to the ZC$m$-$n$ cylinder
(or XC$2m$ cylinder) in the previous literature~\cite{Gong2013}.
All simulations were performed in Julia using GPU acceleration.

\begin{figure*}
    \centering
    \includegraphics[width=16cm]{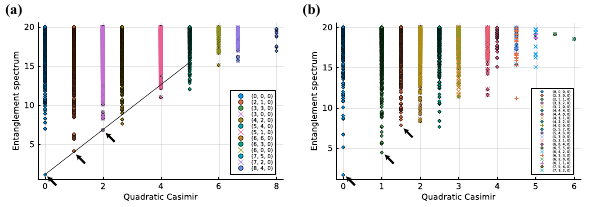}
    \caption{(a) Entanglement spectrum of the $\mathrm{SU}(3)$ Heisenberg model
    on the $12 \times 12$ square lattice with $m=6400$.  For each value,
    an irrep is associated as shown in the legend.  The lowest values for
    (0, 0, 0), (2, 1, 0), (3, 0, 0) and (3, 3, 0) irreps show a linear
    behavior about the quadratic Casimir, as indicated by the black line.
    (b) Entanglement spectrum of the $\mathrm{SU}(4)$ Heisenberg
    model on the honeycomb lattice with $m=12800$.
    ZC6-32 cylinder is used. The entanglement spectrum shows a random
    behavior, which is consistent with the spin liquid ground state.
    Relevant data points are pointed by black arrows.}
    \label{es}
\end{figure*}

For the ground state, we simulate this model up to $m=12800$ for
$L_y=4,\,8,\,10,\,12$, and $m=3200$ for $L_y=6$.  The $L_y=6$
simulation suffers from its gapless nature, and the calculation
cost is most expensive.  The truncation error is close to
machine precision for $L_y=4$, and is less than $10^{-5}$ for $L_y=12$.
From now on, if not specified all the physical quantities are values
after the extrapolation $m \to \infty$.  The extrapolation is done
for energies (resp. entanglement entropies) by an empirical linear
(resp. quadratic) fit with respect to the truncation error with an
error bar being 1/5 of the difference between the extrapolated value
and the value with $m=12800$ ($m=3200$ for $L_y=6$).

The most intriguing feature in the phase diagram (schematically
shown in Fig.~\ref{ov}(a)) about $L_y$ is
the existence of a critical point at $L_y=6$.  A careful analysis
reveals that $L_y=4$ is a rung singlet phase, and $L_y \geq 8$
is a gapped spin liquid phase, while $L_y=6$ is gapless.
This behavior can be understood as a phase transition between
the rung singlet phase and the gapped spin liquid phase, and
its criticality is found to be an $\mathrm{SU}(4)$ level-1
Wess-Zumino-Witten criticality.
Interestingly, this gapless behavior observed on narrow cylinders
is consistent with the Dirac spin-orbital liquid that has been argued
to be unstable in two dimensions~\cite{Calvera2021}.
Our results may provide a direct numerical manifestation of
this theoretical scenario, where the Dirac spin-orbital liquid appears
as a proximate critical state before giving way to a gapped
phase whose available data are most consistent with $Z_4$ topological order.
Details are included in the Supplemental Material~\cite{SM}.

Except for the $L_y=6$ case, the gapped nature of the system
is consistent with its entanglement entropy $S_\mathrm{EE}$ and
spin-spin correlation $\langle P_{ij} \rangle-1/4$.
$S_\mathrm{EE}$ is defined as
$S_\mathrm{EE}=-\mathrm{Tr}\,\rho_A \ln \rho_A$ with
$\rho_A=\mathrm{Tr}_B\,\ket{\Psi}\bra{\Psi}$ for a
bipartition of a system into $A$ and $B$, and $\ket{\Psi}$
is a ground state for the whole system.  As shown
in Fig.~\ref{ov}(b), the entanglement entropy is almost constant for
different entanglement cut.  In addition, in gapped spin liquid phases
the correlation length is around 2-4 sites, which indicates
that $L_y=8,\,10,\,12$ exhibit the same bulk physics
on the accessible cylinders, although larger circumferences are required
to establish the 2D scaling.  We note that when $L_y=4$ a singlet is formed along
the $y$-direction, while this singlet formation is
exponentially suppressed when $L_y>6$.

The absence of $\mathrm{SU}(4)$ symmetry breaking is supported by
the structure of the entanglement spectrum.  It is expected
that the entanglement spectrum has Anderson's tower of states
when the symmetry is broken~\cite{Metlitski2015}.  Anderson's tower of states are
clearly seen in the $\mathrm{SU}(3)$ Heisenberg model on the
square lattice, which is expected to show a three-sublattice
order~\cite{Bauer2012}, as shown in Fig.~\ref{es}(a), while there is no structure
in the $\mathrm{SU}(4)$ Heisenberg model on the honeycomb lattice,
as shown in Fig.~\ref{es}(b).  Indeed, the entanglement spectrum
of the $\mathrm{SU}(4)$ Heisenberg model on the honeycomb lattice
shows no discernible tower-of-states structure,
consistent with an $\mathrm{SU}(4)$-symmetric disordered ground state.

The absence of tetramerization order, which breaks the translation
symmetry, is indicated from the real space structure of
expectation values of bond operators.  Fig.~S2 in Supplemental Material~\cite{SM}
shows the fluctuation of expectation values of bond operators
on the ZC4-12, ZC5-12, and ZC6-12 cylinders.  The fluctuation indeed decays when
$L_y$ gets larger.  The absence of the spontaneous symmetry
breaking of the $\mathrm{SU}(4)$ and translation symmetries
automatically means that the ground states are degenerate when
they are gapped, according to the
Lieb-Schultz-Mattis-Affleck-Yamada-Oshikawa-Jackeli theorem~\cite{Yamada2018,Yamada2021,Lieb1961,Affleck1986}.
In this sense, we numerically inferred the existence of
topological order, \textit{i.e.} gapped spin liquid ground
states, in the $\mathrm{SU}(4)$ Heisenberg model on the honeycomb
lattice.

\begin{figure*}
    \centering
    \includegraphics[width=16cm]{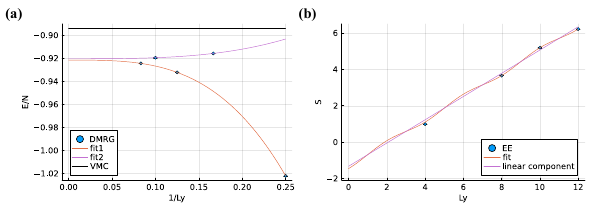}
    \caption{(a) Finite size scaling of energy about $L_y$. Blue dots are
    plotted for each $L_y$.  A red line is for $L_y \mod 4 = 0$, and
    a purple line is for $L_y \mod 4 = 2$.  A black line shows the VMC
    result from Ref.~\cite{Corboz2012}.
    (b) Finite size scaling of entanglement entropy about $L_y$.
    Blue dots are plotted for each $L_y$, while the point for $L_y=6$ is
    missing because of its gapless nature.  A red line is for an exact
    fit for $L_y>6$, and a purple line shows its linear component.}
    \label{fss}
\end{figure*}

\textit{2D limit}. ---
From now on, all data points are plotted after the extrapolation
about $L_x \to \infty$.  For energies (resp. entanglement entropies),
a linear (resp. exponential) fit about $1/\Delta L_x$ (resp. $L_x$) is
used in the extrapolation with $\Delta L_x=L_x-12$,
where $L_x=12$ is set to be a reference to extract the boundary effect.  We computed until $L_x=32$ ($L_x=36$
for $L_y=6$), and checked that the results are well-converged
about $L_x$.  Then, as shown in Fig.~\ref{fss}(a), the energy per
site in the thermodynamic limit is extracted from the power law
fitting about $L_y$, \textit{i.e.} $E/N=qL_y^{-p} + r$.  The best
fit is achieved by $p=3.235$ and $E/N$ in the 2D limit is
estimated as,
\begin{equation}
    E_\mathrm{2D}/N=-0.9210(6).\label{e2d}
\end{equation}
This value is significantly lower than the previous estimate $-0.894$ from variational
Monte Carlo (VMC) simulations~\cite{Corboz2012}.  Even if the estimation
(Eq.~\eqref{e2d}) is not rigorous, all the finite-size points plotted in Fig.~\ref{fss}(a)
are well below the VMC result (shown in a black line), which strongly disfavors
the previously proposed $\pi$-flux VMC ansatz as a quantitative description
of the ground state.  While we admit that only three points
in the gapped spin liquid phase ($L_y=8,\,10,\,12$) are used, the systematic extrapolation
will require additional calculations with a larger $L_y$ or a different boundary condition,
which will be left for future work.
We note that the previous work based on DMRG without non-Abelian
symmetry implementation achieved much smaller bond dimensions and
system sizes~\cite{Jin2023}.

The topological entanglement entropy $\gamma_\mathrm{top}$ is
estimated only from the data of $L_y=8,\,10,\,12$, as shown
in Fig.~\ref{fss}(b).  We note that in Fig.~\ref{fss}(b) the point
for $L_y=6$ is missing because of the gapless nature,
and we note that for the $L_y=10$ case, where
the even-odd effect appears, we always use the system with
a multiple of 4 sites in both sides of the central cut,
where both sides admit a singlet formation in accordance with
the Lieb-Schultz-Mattis-type theorem.  This is
a little tricky because there is a strong ``mod 4'' effect.
Because the tendency is different between $L_y=8,\,12$ and
$L_y=10$, we additionally include the oscillatory term
in addition to the linear fit.
\begin{equation}
    S_\mathrm{EE}=a+bL_y+c\cos\left(\frac{\pi}{2} L_y\right).
\end{equation}
From this we can extract $\gamma_\mathrm{top}=-a$.  An exact fit
is achieved by $\gamma_\mathrm{top}=1.33(3)$, which is very
close to $\ln(4)=1.386$.
The fitted value is consistent with $Z_4$ topological order,
which would imply a 16-fold torus ground-state degeneracy,
although the limited number of circumferences and the sensitivity to
the fitting form prevent a definitive identification.
While the estimation here is not rigorous, the Lieb-Schultz-Mattis-type
theorem rules out a unique short-range-entangled ground state and therefore supports intrinsic topological order
even without the information of the topological entanglement entropy,
under the assumptions of a gap and unbroken $\mathrm{SU}(4)$ and translation symmetries~\cite{Yamada2022}.

\textit{Summary}. ---
From the combination of (i) the highly accurate variational DMRG energy
well below the Dirac spin liquid ansatz, (ii) the absence of both
$\mathrm{SU}(4)$ and lattice symmetry breaking as revealed by entanglement
spectra and bond observables, and (iii) a finite topological entanglement
entropy close to $\ln(4)$, we conclude that the ground state of
the $\mathrm{SU}(4)$ Heisenberg model on the honeycomb
lattice is most consistent with a gapped $Z_4$ spin liquid.  This is a natural
generalization of Anderson's resonating valence bond state to $\mathrm{SU}(4)$~\cite{Anderson1973}.
The previously reported gapless Dirac spin liquid~\cite{Corboz2012}
may be related to the $L_y=6$ gapless state, and it is possible that
this model may appear gapless at small system sizes, in remarkable agreement
with the field-theoretical scenario proposed by Calvera and Wang~\cite{Calvera2021}.
It is an intriguing open direction to
investigate the fermionic version of our model because an exotic
superconductivity like a charge-4e superconductivity is expected
upon doping~\cite{Anderson1987}.

The possible relevance of the present results to experimental proposals
and studies can be viewed from two complementary perspectives.
For solid-state spin-orbital materials~\cite{Yamada2018,Yamada2021},
the $\mathrm{SU}(4)$-symmetric model should be regarded as a parent Hamiltonian
for a proximate topological phase: although realistic materials inevitably
contain $\mathrm{SU}(4)$-breaking perturbations, the gapped spin-orbital liquid
found here provides characteristic signatures, such as the absence of conventional
$\mathrm{SU}(4)$ symmetry breaking, a finite correlation length, and
possible broad excitation continua expected from fractionalization,
that can guide future neutron-scattering and
related spectroscopic studies of candidate materials such as $\alpha$-ZrCl$_3$.
For cold atomic systems with four internal states~\cite{Cazalilla2014},
our results establish the $\mathrm{SU}(4)$ Heisenberg model on the honeycomb lattice
as a concrete target Hamiltonian for quantum simulation platforms,
where long-range entanglement, rather than conventional symmetry breaking,
stabilizes an exotic quantum phase of matter.  This provides a concrete route
by which spin-orbital exchange interactions can generate intrinsic topological order
in a microscopic 2D Hamiltonian.

\begin{acknowledgments}
We thank K.~Penc and F.~Pollmann for insightful discussions.
We also thank H.-K.~Jin and J.~Knolle for informative advice.
The computation in this work has been done using the facilities of the Supercomputer Center,
the Institute for Solid State Physics, the University of Tokyo.
M.G.Y. is supported by JST PRESTO Grant No. JPMJPR225B.
M.G.Y. is partly supported by Multidisciplinary Research Laboratory System
for Future Developments, Osaka University, and by JSPS KAKENHI Grant
Nos. JP22K14005 and JP26K17088.
This work is a part of the outcome of research performed
under a Waseda University Grant for Special Research Projects
(Project number: 2026C-662).
\end{acknowledgments}

\bibliography{paper}

\end{document}


\title{Supplemental Material for \\ ``Gapped topological spin-orbital liquid on the honeycomb lattice''}

\author{Masahiko G. Yamada}
\affiliation{Waseda Institute for Advanced Study, Waseda University, Shinjuku-ku, Tokyo 169-0051, Japan.}
\affiliation{Department of Physics, School of Science, the University of Tokyo, Hongo, Bunkyo-ku 113-0033, Japan.}

\maketitle

\onecolumngrid

\appendix

\section{I. Density Matrix Renormalization Group}

Our density matrix renormalization group (DMRG) algorithm is deeply inspired
by the pioneering work by Nataf and Mila~\cite{Nataf2018}.
A part of the code is influenced by Simple DMRG~\cite{simple-dmrg}.

In order to improve the performance, we made an important modification
to the algorithm by Nataf and Mila.  We never keep the wavefunction in
the vector form, but rather we keep it in matrix form.
The matrix is found to be very sparse, and in the ground state sector
nonzero values appear only in rectangles where two irreducible representations (irreps)
combine into the ground state representation.  Thus, it is enough
to keep each matrix which represents a part of the ground state, which drastically
speeds up the Lanczos iteration.  Details will be discussed in the future
publication.

As described in the main text, $9\nu$ coefficients are computed
from subduction coefficients (SDCs) of symmetric groups.  We use a different
gauge in the definition of SDCs from the previous research~\cite{Nataf2018}.
Our gauge fixing scheme will be discussed in the future publication.  The calculation
of SDCs is by the standard procedure based on standard Young tableaux (SYTx)~\cite{Chen2002}.
We employ the inverse Wilf-Rao-Shankar method to index and retrieve
SYTx~\cite{Wilf1977,Rao2015}.  Details will be discussed in the future publication.

We note that there is a different approach called QSpace to this
problem~\cite{Weichselbaum2024}.  QSpace is not as fast as our method
because the calculation still relies on the structure of Clebsch-Gordan
coefficients (CGCs).  We do not need any explicit computation of CGCs.

\section{II. Construction of the Heisenberg interaction}

The following formulas are used to construct the $\mathrm{SU}(N_c)$ Heisenberg interaction,
where $N_c$ is the number of colors.  All of them are necessary to create Hamiltonians
in DMRG simulations.

In the case of the Heisenberg interaction inside the same block, we need the following type.
\begin{equation}
P_{ij}^{\alpha\beta}-\frac{1}{N_c} \to
\frac{(-1)^{N_c + 1}(N_c^2-1)^{3/2}}{N_c} \sum_{\gamma,\gamma'}
\begin{bmatrix}
\alpha & \ydiagram{1} & \gamma \\
\cdot & \ydiagram{2,1,1} & \ydiagram{2,1,1} \\
\alpha & \ydiagram{1} & \gamma'
\end{bmatrix}
\begin{bmatrix}
\gamma & \ydiagram{1} & \beta \\
\ydiagram{2,1,1} & \ydiagram{2,1,1} & \cdot \\
\gamma' & \ydiagram{1} & \beta
\end{bmatrix},
\end{equation}
where $\alpha$ and $\beta$ are specific irreps,
which actually select the sector of the matrix element of $P_{ij}$ belonging
to the $\alpha\beta$ sector of the operator, $\gamma$ and $\gamma'$ run over
all relevant irreps, and $\cdot$ represents the trivial representation.

In the case of the superblock Hamiltonian in the ground state sector of $\gamma$,
we need the following type.
\begin{equation}
P_{ij}^{\alpha\beta(\gamma)}-\frac{1}{N_c} \to
\frac{(-1)^{N_c + 1}(N_c^2-1)^{3/2}}{N_c} \sum_{\delta,\delta',\epsilon,\epsilon'}
\begin{bmatrix}
\alpha & \ydiagram{1} & \delta \\
\cdot & \ydiagram{2,1,1} & \ydiagram{2,1,1} \\
\alpha & \ydiagram{1} & \delta'
\end{bmatrix}
\begin{bmatrix}
\beta & \ydiagram{1} & \epsilon \\
\cdot & \ydiagram{2,1,1} & \ydiagram{2,1,1} \\
\beta & \ydiagram{1} & \epsilon'
\end{bmatrix}
\begin{bmatrix}
\delta & \epsilon & \gamma \\
\ydiagram{2,1,1} & \ydiagram{2,1,1} & \cdot \\
\delta' & \epsilon' & \gamma
\end{bmatrix},
\end{equation}
where $\delta$, $\delta'$, $\epsilon$, and $\epsilon'$ run over all relevant irreps.

Therefore, we can conclude that the following four types are sufficient to construct
the $\mathrm{SU}(4)$ Heisenberg interaction used in the main text. We note that the
first one is only used to expand the left and right blocks without an additional factor.
\begin{equation}
\begin{bmatrix}
\alpha & \ydiagram{1} & \beta \\
\ydiagram{2,1,1} & \cdot & \ydiagram{2,1,1} \\
\alpha' & \ydiagram{1} & \beta'
\end{bmatrix}
\begin{bmatrix}
\alpha & \ydiagram{1} & \beta \\
\cdot & \ydiagram{2,1,1} & \ydiagram{2,1,1} \\
\alpha & \ydiagram{1} & \beta'
\end{bmatrix}
\begin{bmatrix}
\alpha & \ydiagram{1} & \beta \\
\ydiagram{2,1,1} & \ydiagram{2,1,1} & \cdot \\
\alpha' & \ydiagram{1} & \beta
\end{bmatrix}
\begin{bmatrix}
\alpha & \beta & \gamma \\
\ydiagram{2,1,1} & \ydiagram{2,1,1} & \cdot \\
\alpha' & \beta' & \gamma
\end{bmatrix}.
\end{equation}
We note that in order to show the adjoint representation $\ydiagram{2,1,1}$ simply
it was shown assuming $N_c=4$ here, but the same formulas are valid for general $N_c$.

\section{III. Critical phase at $L_y=6$}\label{cft}

Gapless nature of the critical phase at $L_y=6$ is first found from the fact that
the convergence of DMRG is very difficult compared to other gapped phases.  Indeed,
$m=3200$ is the largest bond dimension we can reach for $L_y=6$ cylinders, while
we struggle to obtain a well-converged ground state.

\begin{figure}
\centering
\includegraphics[width=18cm]{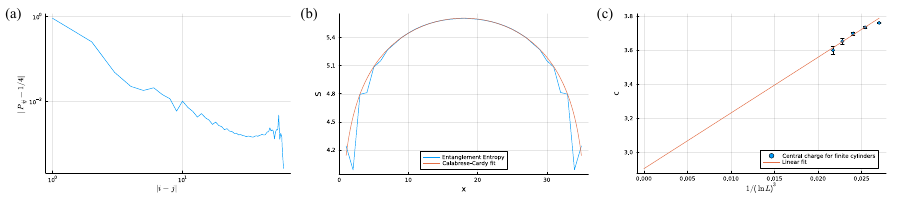}
\caption{(a) Spin-spin correlation function for the $L_y=6$ cylinder with $L_x=36$ and its nearly power-law decay.
(b) Entanglement entropy for the $L_y=6$ cylinder with $L_x=36$ and its fitting by the Calabrese-Cardy formula.
(c) Ziman-Schulz finite size scaling to estimate the central charge.}
\label{f}
\end{figure}

In addition, the spin-spin correlation function shows a nearly power-law decay
as shown in Fig.~\ref{f}(a). The entanglement entropy is well fitted by
the following Calabrese-Cardy formula~\cite{Calabrese2009}.
\begin{equation}
S(x)=\frac{c}{6}\log\left[\frac{2L}{\pi}\sin\left(\frac{\pi x}{L}\right)\right].
\end{equation}
As an example, we show an entanglement entropy for the $L_x=36$ cylinder in Fig.~\ref{f}(b).
It is clear that the entanglement entropy is well fitted by the Calabrese-Cardy
formula except for the small even-odd effect, and the same is true for any $L_x$.
The finite size scaling of the entanglement entropy is based on Ziman and Schulz~\cite{Ziman1987},
as shown in Fig.~\ref{f}(c), and from that we estimate the central charge as $c=2.90(11)$,
which is consistent with an $\mathrm{SU}(4)$ level-1 Wess-Zumino-Witten criticality with $c=3$,
as explained in the main text.
The gapless behavior observed on $L_y=6$ cylinders should not be interpreted as
a stable two-dimensional phase.  Instead, its central charge $c=3$ and conformal structure
strongly suggest that it represents a quasi-one-dimensional remnant of the Dirac
spin-orbital liquid, which acts as a parent phase proximate to the gapped phase
that is most consistent with $Z_4$ topological order in the available
two-dimensional extrapolation, as the Dirac spin-orbital liquid is known
to be unstable in the two-dimensional limit~\cite{Calvera2021}.

While the results obtained to estimate the central charge are not very accurate
due to the limited bond dimension and the slow convergence, the gapless nature
of the $L_y=6$ phase is obvious.  Thus, we can conclude that the critical transition
indeed occurs at $L_y=6$.

\section{IV. Fluctuation of expectation values of bond operators}

\begin{figure}
    \centering
    \includegraphics[width=12cm]{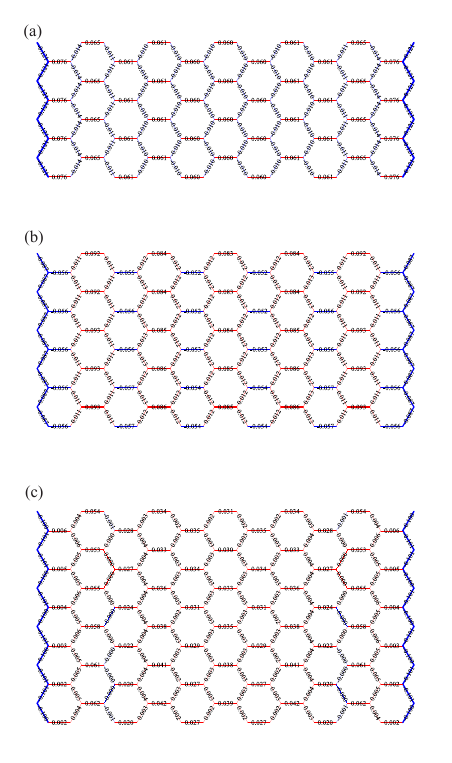}
    \caption{Fluctuation of expectation values of bond operators
    is displayed by the thickness of bonds for the (a) ZC4-12,
    (b) ZC5-12, and (c) ZC6-12 cylinders.  Blue bonds indicate
    a minus value and red bonds indicate a plus value with respect
    to the average, and the exact value is annotated on the bond.}
    \label{pij}
\end{figure}

The bond-observable data show no clear signature of
tetramerization within the accessible cylinder sizes.
Fig.~\ref{pij} shows their evolution
on the (a) ZC4-12, (b) ZC5-12, and (c) ZC6-12 cylinders.
In particular, Fig.~\ref{pij}(c) shows that the bond fluctuations
are suppressed in the bulk, with no signature of translation symmetry breaking.
The boundary effects decay sufficiently toward the center of the cylinder,
confirming that the bulk properties are unchanged for a larger $L_x$.
Only the $L_y = 10$ data are affected by the even-odd effect,
in accordance with the Lieb-Schultz-Mattis-type constraint.
Finally, we note that our presentation of the expectation values
of bond operators follows a previous DMRG study of the spin-1/2
$J_1$-$J_2$ Heisenberg model on the honeycomb lattice~\cite{Gong2013}.

\section{V. Convergence about $L_x$ and correlation data}

\begin{figure}
    \centering
    \includegraphics[width=12cm]{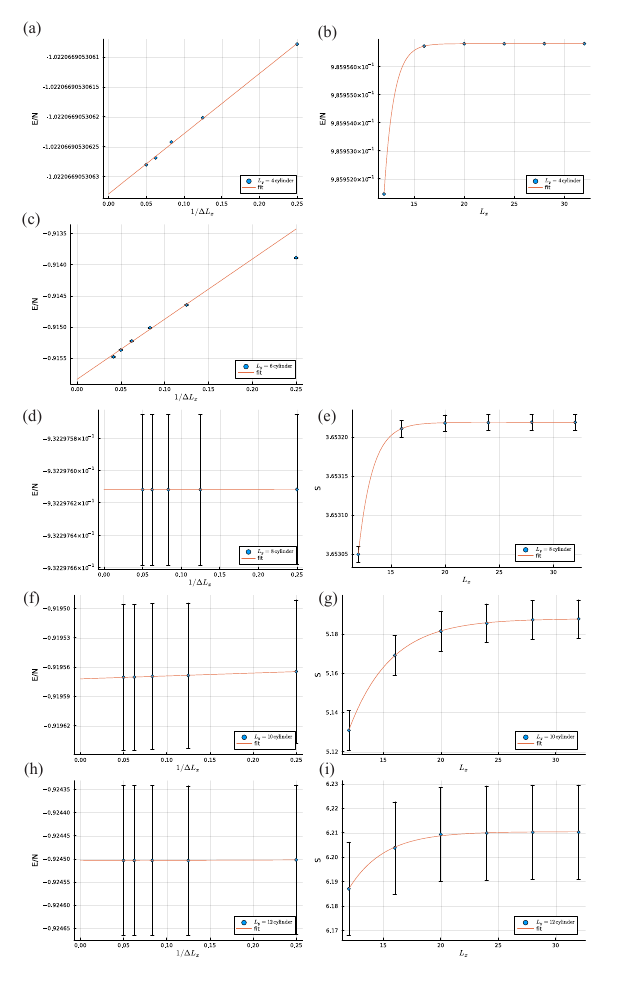}
    \caption{(a) Energy convergence about $L_x$ for the $L_y=4$ cylinder.
    For the rung singlet phase of $L_y=4$, the truncation error is negligibly small
    and the errorbar is omitted.
    (b) Entanglement entropy convergence about $L_x$ for the $L_y=4$ cylinder.
    Again for the rung singlet phase of $L_y=4$, the truncation error is negligibly small
    and the errorbar is omitted.
    (c) Energy convergence about $L_x$ for the $L_y=6$ cylinder.
    We note that the entanglement entropy for $L_y=6$ is omitted and the detailed analysis is included in Section~III.
    (d) Energy convergence about $L_x$ for the $L_y=8$ cylinder.
    (e) Entanglement entropy convergence about $L_x$ for the $L_y=8$ cylinder.
    (f) Energy convergence about $L_x$ for the $L_y=10$ cylinder.
    (g) Entanglement entropy convergence about $L_x$ for the $L_y=10$ cylinder.
    (h) Energy convergence about $L_x$ for the $L_y=12$ cylinder.
    (i) Entanglement entropy convergence about $L_x$ for the $L_y=12$ cylinder.}
    \label{conv}
\end{figure}

In order to clarify the convergence of the DMRG observables
with respect to $L_x$, we show the extrapolation of
the ground state energy and the entanglement entropy about
the bipartition at the center of the cylinder.  Energies
are fitted by a linear function of $1/\Delta L_x$ and
entanglement entropies are fitted by an exponential
function of $L_x$.  In the finite-DMRG calculations,
in order to minimize the boundary effects, we use the $L_x=12$ data
as the reference.  We use the $L_x=12$ cylinder because the
suppression of the boundary effect at the center of the cylinder is
clear in Fig.~\ref{pij}.  For systems with $L_x \geq 16$,
the corresponding quantities are obtained by subtracting
the $L_x=12$ values, and the resulting differences
are plotted as a function of $\Delta L_x=L_x-12$.
Fig.~\ref{conv} shows the convergence of energies and entanglement entropies
for different $L_y$ cylinders.  In Fig.~\ref{conv}, (a), (c), (d), (f), and (h)
are for energies with $L_y=4$, $6$, $8$, $10$, and $12$, respectively, and
(b), (e), (g), and (i) are for entanglement entropies with $L_y=4$, $8$, $10$, and $12$,
respectively.  Here the entanglement entropy for the gapless case
of $L_y=6$ is omitted.

It is clear that the errorbar of energy is much larger than the
difference about $L_x$, where the errorbar is estimated
from the truncation error of DMRG calculations, for gapped
cases.  Except for the critical case of $L_y=6$, the convergence
about the cylinder length $L_x$ is confirmed for energies.
As for the entanglement entropy, the exponential convergence
is confirmed for gapped cases, at least for $L_x \gtrsim 20$.
As for the $L_x$ dependence for the case of $L_y=6$, Section~III
gives a detailed analysis based on the critical scaling of
the conformal field theory.

\begin{figure}
    \centering
    \includegraphics[width=18cm]{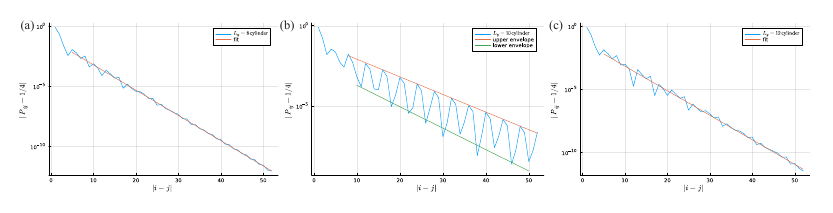}
    \caption{(a) Spin-spin correlation function for the $L_y=8$ cylinder with $L_x=32$ and its exponential decay.
    (b) Spin-spin correlation function for the $L_y=10$ cylinder with $L_x=32$ and its exponential decay.
    An oscillatory behavior is observed, and the exponential fit is done separately for the upper and lower envelopes.
    (c) Spin-spin correlation function for the $L_y=12$ cylinder with $L_x=32$ and its exponential decay.}
    \label{sisj}
\end{figure}

Finally, we quickly show the spin-spin correlation analysis.
The spin-spin correlation $\langle P_{ij} \rangle-1/4$ is
evaluated along an armchair chain along the $x$-direction.
In Fig.~\ref{sisj}, we display the exponential decay of
the spin-spin correlation function for gapped spin liquid phases
of $L_y=8$, $10$, and $12$.  As a function of the distance in the
$x$ direction, the correlation function is fitted by an exponential function,
and the correlation length is estimated as 2.07(2), and 2.23(4)
for $L_y=8$, and $12$ cylinders, respectively.  As clearly seen in Fig.~\ref{sisj}(b),
the $L_y=10$ cylinder shows an oscillatory behavior, and the exponential fit is
done separately for the upper and lower envelopes. The correlation length is estimated as
3.96(4) for the upper envelope, and 3.26(26) for the lower envelope.

\bibliography{suppl}